\documentclass{article}

\usepackage{ijcai13}
\usepackage{amssymb}
\usepackage[dvips]{graphicx}
\usepackage{times}
\newtheorem{Theorem}{Theorem}
\newtheorem{Lemma}{Lemma}

\newtheorem{Example}{Example}

\title{Allocating Indivisible Resources under Price Rigidities in Polynomial Time}
\author{Wei Huang$^{1}$, Jian Lou$^{1}$, \and Zhonghua Wen$^{2}$\\
1. School of Computer Science and Technology, USTC, Hefei, China\\
huangbodao@gmail.com, loujian@mail.ustc.edu.cn\\
2. College of Information Engineering, Xiangtan
University, Xiangtan, China\\
zhonghua@xtu.edu.cn}

\begin{document}

\maketitle

\begin{abstract}
In many realistic problems of allocating resources, economy
efficiency must be taken into consideration together with social
equality, and price rigidities are often made according to some economic and social
needs. We study the computational issues of dynamic mechanisms for selling multiple
indivisible items under price rigidities. We propose a polynomial algorithm
that can be used to find over-demanded sets of items, and then introduce a dynamic
mechanism with rationing to discover constrained Walrasian equilibria under price
rigidities in polynomial time. We also address the computation of sellers' expected profits and items'
expected prices, and discuss strategical issues in the
sense of expected profits.

\end{abstract}

\section{Introduction}

Problem of allocating resources among selfish agents has been a
well-established research theme in economics and recently becomes
an emerging research topic in AI because AI methodologies can
provide computational techniques \cite{Rothkopf,Sandholm,Huang} to
the balancing of computation tractability and economic (or
societal) needs in these problems.

Dynamic mechanisms for resource allocation are trading mechanisms
for discovering market-clearing prices and equilibrium allocations
based on price adjustment processes \cite{Ausubel,Gul1,Huang}
Assume a seller wishes to sell a set of indivisible items to a
number of buyers. The seller announces the current prices of the
items and the buyers respond by reporting the set of items they
wish to buy at the given prices. The seller then calculates the
over-demanded set of items and increases the prices of
over-demanded items. This iterative process continues until all
the selling items can be sold at the prices at which each buyer is
assigned with items that maximize her personal net benefit.

Different from one-shot combinatorial auctions\cite{Cramton}, the main issue of
a dynamic mechanism is whether the procedure can lead to an
equilibrium state (Walrasian equilibrium) at which all the selling
items are effectively allocated to the buyers (equilibrium
allocation) and the price of items gives the buyers their best
values \cite{Gul2,Kelso,Lehmann,Sun}.

Most of the discussions on the issues of dynamic mechanisms are
based on market models in which there does not exist price
rigidities. In fact, ``good" allocations must look after both
sides economy efficiency and social equality, and price rigidities
may play a key role in some of these problems. For instance, in an
estate bubble period, housing cost is unbearable for most of the
members of society. The government may need to allocate some
housing resources (whose prices are not completely flexible but
restricted under some price rigidities) to middle-income earners.
On one hand, the lower bound prices can be made according to some
basic economic requirements (e.g., construction costs); on the
other hand, the upper bound prices \footnote{Note that since upper
bound prices are often set for the sake of equality between social
members (who have some but limited pay ability), they generally
accompany a limit to the number of resources one member can get.}
should be made according to some realistic social foundation
(e.g., average income level or pay ability). It is well-known that
a Walarasian equilibrium exists in the economy when there are no
price rigidities. In the case of price restrictions, a Walrasian
equilibrium may not exist since the equilibrium price vector may
not be admissible.

Talman and Yang studied the equilibrium allocation of heterogeneous
indivisible items under price rigidities, and proposed the
concept of constrained Walrasian equilibria \cite{Talman}. A
constrained Walrasian equilibrium consists of a price vector
$\textbf{p}$, a rationing system $R$, and a (constrained)
equilibrium allocation $\pi$ \cite{Lehmann} s.t. $\textbf{p}$
obeys the price rigidities, and $\pi$ assigns each buyer an
item (permitted by $R$) that maximizes her personal net
benefit at $\textbf{p}$. They also proposed two dynamic
auction procedures that produce constrained
Walrasian equilibria. However, the computational issues of these procedures have not been touched.

In this paper, we present a polynomial algorithm that
can be used to find over-demanded sets of items, and then
introduce a dynamic mechanism(called MAPR) with rationing to discover
constrained Walrasian equilibria under price rigidities in
polynomial time. In
MAPR, buyers compete with each other (with the help of the
seller) on prices of items for multiple rounds. In each
round, the seller announces the current price vector
(initially, the lower bound price vector) of the items that
remain, then the buyers respond by reporting the set of resources
they wish to buy, then the seller computes a minimal
over-demanded set $X_{min}$ of the items. If
$X_{min}=\emptyset$ then the final allocation is computed by the
RM subroutine and MAPR stops. Otherwise if all the prices of the
items in $X_{min}$ are less than their upper bounds then the
seller increases them; else an item $a\in X_{min}$ (whose
price is on its upper bound) is picked and the buyers who only
demand some items (including $a$) in $X_{min}$ draw lots for
the right to buy $a$. Since MAPR's execution process is
nondeterministic, we define the concepts of buyers' expected
profits and items' expected prices, and consider strategical
issues (in the sense of expected profit) in MAPR.

Here are main contributions of our work:
\begin{itemize}
 \item We address the computational problems of dynamic auction proposed by [Talman and Yang, 2008], where these problems have not been touched.
 \item \cite{Talman} has not finished the proof about the existence of constrained Walrasian equilibrium. We propose an algorithm to get the final allocation and several lemmas to prove the criteria required in constrained Walrasian equilibrium.
 \item We defined the ``expected profits'' and ``expected prices'' and discuss strategical issues.
\end{itemize}

This paper is structured as follows. First, we review some basic
notions that are relevant to our work (see \cite{Talman} for
further details and examples). Second, we represent demand
situations with bipartite graphs. Third, we address the
computation of minimal over--demanded sets of items. Fourth,
we present MAPR, and prove formally that it yields a constrained
Walrasian equilibrium in polynomial time. Fifth, we consider
strategical issues in MAPR. Finally, we draw some conclusions.

\section{Preliminaries}

Consider a market situation where a seller wishes to sell a finite set $X$ of
indivisible items to a finite
number of buyers
$N=\{1,2,\ldots,n\}$. The item $o\in X$ is a dummy item
which can be assigned to more than one buyer. Items (eg.,
houses or apartments) in $X\setminus\{o\}$ may be heterogeneous.

A price vector $\textbf{p}\in\mathbb{Z}_+^X$ assigns a
non-negative integer to each $a\in X$ and $\textbf{p}_a$ is the
price of $a$ under $\textbf{p}$. It is required that
$\textbf{p}_a$ is not completely flexible and restricted to an
interval $[\underline{\textbf{p}}_a, \overline{\textbf{p}}_a]$
s.t.
$\underline{\textbf{p}}_a,\overline{\textbf{p}}_a\in\mathbb{Z}_+$,
$\underline{\textbf{p}}_a\leq\overline{\textbf{p}}_a$, and
$0=\underline{\textbf{p}}_o=\overline{\textbf{p}}_o$. We say
$\underline{\textbf{p}}$ and $\overline{\textbf{p}}$ as the lower
and upper bound price vectors. $P=\{\textbf{p}\in
\mathbb{Z}_+^X|(\forall a\in
X)\underline{\textbf{p}}_a\leq\textbf{p}_a\leq\overline{\textbf{p}}_a\}$
is called the set of \emph{admissible} price vectors. Each $i\in
N$ has an integer value function, i.e., $u_i : X \rightarrow
\mathbb{Z}_+$. $u_i(a)$ is i's valuation to item $a$. We
assume $u_i$ is i's private information, $u_i(o) = 0$, and i can
pay $\max_{a\in X}\overline{\textbf{p}}_a$ units of money. We say
$E=\langle N,X,\{u_i\}_{i\in N}\rangle$ is an \textit{economy}.

A rationing system is a function $R:N\times X\rightarrow\{0,1\}$
s.t. $R(i,o)=1$ for every $i\in N$. $R(i,a)=1$ means that buyer
$i$ is allowed to demand item $a$, while $R(i,a)=0$ means that
$i$ is not allowed to demand $a$. At $\textbf{p}\in P$ and
rationing system $R$, the indirect utility $V_i(\textbf{p},R)$ and
constrained demand $D_i(\textbf{p},R)$ of buyer $i$ is given by:
$V_i(\textbf{p},R)=\max\{u_i(a)-\textbf{p}_a|a\in X\textrm{ and
}R(i,a)=1\}$, and $D_i(\textbf{p},R)=\{a\in X|R(i,a)=1\textrm{ and
}u_i(a)-\textbf{p}_a=V_i(\textbf{p},R)\}$. An \textit{allocation}
of $X$ is a function $\pi:N\rightarrow X$ s.t. $\pi(i)\neq \pi(j)$
if $j\neq i$ and $\pi(i)\in X\setminus\{o\}$. $\pi$ is an
\textit{equilibrium allocation} if $\pi(i)\in D_i(\textbf{p},R)$
for all $i\in N$.

$\langle \textbf{p},R,\pi\rangle$ is a \textit{constrained
Walrasian equilibrium} if \textbf{(1)} $\textbf{p}\in P$, $R$ is a
rationing system, \textbf{(2)} $\pi$ is an equilibrium allocation,
\textbf{(3)} $\textbf{p}_a=\underline{\textbf{p}}_a$ if
$\pi(i)\neq a$ for all $i\in N$, \textbf{(4)}
$\textbf{p}_a=\overline{\textbf{p}}_a$ and $\pi(i)=a$ for some
$i\in N$ if $R(j,a)=0$ for some $j\in N$, and \textbf{(5)} $a\in
D_i(\textbf{p},R')$ if $R(i,a)=0$, where $R'(j,b)=R(j,b)$ for all
$\langle j,b\rangle\in N\times X$ except $R'(i,a)=1$.

Conditions \textbf{(1)} and \textbf{(2)} need no explanation.
Condition \textbf{(3)} says that if the price of a item is greater
than its lower bound then it must be assigned to some buyer.
\textbf{(4)} states that if an buyer is not allowed to demand some
items then the item must be assigned to another buyer at its upper
bound price. Condition \textbf{(5)} says that if an buyer is
allowed to demand a item which she was not allowed to demand, then
she will demand the item. To sum up, \textit{constrained Walrasian
equilibrium} is a equilibrium state under price rigidities. All
the five conditions make a balance between efficiency and
equality.

The following example is modified from the one given in
\cite{Talman}. It illustrates the notions introduced in this
section and will be used throughout the paper.
\begin{Example}
\begin{small}
Let $E=\langle N,X,\{u_i\}_{i\in N}\rangle$ be an economy such
that $N=\{1,2,3,4,5\}$, $X=\{o,a,b,c,d\}$, and buyers' values are
given in Table 1; price vector $\textbf{p}=(0,5,4,4,7)$; and $\pi$
be an allocation of X such that $\pi(1)=o$, $\pi(2)=c$,
$\pi(3)=b$, $\pi(4)=a$, and $\pi(5)=d$. Suppose the lower and
upper bound price vectors are $\underline{\textbf{p}}=(0, 5, 4, 1,
5)$, and $\overline{\textbf{p}}=(0, 6, 6, 4, 7)$, respectively. So
$\textbf{p}$ is an admissible price vector. Let $R$ be a rationing
system such that $R(i,x)=1$ for all $\langle i,x\rangle\in N\times
X$ except that $R(3,c)=R(1,c)=0$. For each buyer $i\in N$,
$V_i(\textbf{p},R)$ and $D_i(\textbf{p},R)$ are also shown in
Table 1. Obviously, $\langle\textbf{p},R,\pi\rangle$ is a
constrained Walrasian equilibrium.
\end{small}
\end{Example}
\begin{table}
\caption{Values, Indirect Utilities, and Constrained Demand} \setlength\tabcolsep{3pt}
\begin{center}
\begin{footnotesize}
\begin{tabular}{c|c|c|c|c|c|c|c}
\hline
buyer i & $u_{i}(o)$ & $u_{i}(a)$ & $u_{i}(b)$ & $u_{i}(c)$ & $u_{i}(d)$ & $V_i(\textbf{p},R)$ & $D_i(\textbf{p},R)$\\
\hline
1 & 0 & 4 & 3 & 5 & 7 & 0 & \{o,d\}\\
2 & 0 & 7 & 6 & 8 & 3 & 4 & \{c\}\\
3 & 0 & 5 & 5 & 8 & 7 & 1 & \{b\}\\
4 & 0 & 9 & 4 & 3 & 2 & 4 & \{a\}\\
5 & 0 & 6 & 2 & 4 & 10 & 3 & \{d\}\\
\hline
\end{tabular}
\end{footnotesize}
\end{center}
\end{table}

\section{Demand Situation and Maximum Consistent Allocation}

Given an economy $E=\langle N,X,\{u_i\}_{i\in N}\rangle$, we call
$\mathcal{D} = (D_i)_{i\in N}$ a \textit{demand situation} of $E$
if there is a price vector $\textbf{p}$ and a rationing system $R$
such that $D_i = D_i(\textbf{p},R)$ for all $i\in N$. An
allocation $\pi$ is \emph{consistent} with $\mathcal{D}$ if
$\pi(i)\in D_i\cup\{o\}$ for all $i\in N$. $\pi$ is maximum if
$|\{i\in N|o\not\in D_i\; and\;\pi(i)\neq o\}|$ $\geq|\{i\in
N|o\not\in D_i\; and\;\pi'(i)\neq o\}|$ for every allocation
$\pi'$ consistent with $\mathcal{D}$.

$\mathcal{D}$ can be represented as a bipartite graph
$\textsc{BG}(\mathcal{D}) =\langle N'\cup X',\mathcal{E}\rangle$
where $N'=\{i\in N|o\not\in D_i\}$, $X'=\bigcup_{i\in N'}D_i$, and
$\mathcal{E}=\{\{i,a\}|i\in N',a\in D_i\}$. A \textit{matching} in
$\textsc{BG}(\mathcal{D})$ is a subset $M$ of $\mathcal{E}$ s.t.
$e\cap e'=\emptyset$ for all $e,e'\in M$ with $e\neq e'$. $M$ is
maximum if $|M'|\leq|M|$ for each matching $M'$.

It is not hard to see that a matching $M$ in
$\textsc{BG}(\mathcal{D})$ determines an allocation consistent
with $\mathcal{D}$. $\pi^M$ denotes the allocation determined by
$M$, that is, $\pi^M(i)=a$ if $\exists\{i,a\}\in M$, and
$\pi^M(i)=o$ otherwise. Suppose $M$ is maximum, then $\pi^M$ is
maximum and it is easy to find that: there exists an equilibrium
allocation $\Leftrightarrow$ $|M|=|\{i\in N|o\not\in D_i\}|$
$\Leftrightarrow$ $\pi^M$ is an equilibrium allocation.

In fact, to find a maximum matching in a bipartite graph is a pure
combinatorial optimization problem, which can be addressed in
polynomial time. \cite{Schrijver} presents the matching augmenting
algorithm $\textsc{Ma}$, which takes a bipartite graph
$\mathcal{G}=\langle \mathcal{V},\mathcal{E}\rangle$ and a
matching $M$ in $\mathcal{G}$ as input, and outputs a matching
$\textsc{Ma}(\mathcal{G},M)=M'$ s.t. $|M'|\geq|M|$ and
$\bigcup_{e\in M'}e\supseteq \bigcup_{e\in M}e$ in time
$O(|\mathcal{E}|)$. So a maximum matching can be found in time
$O(|\mathcal{V}||\mathcal{E}|)$ (as we do at most $|\mathcal{V}|$
iterations), i.e., $O(|N||X|\min(|N|,|X|))$. In the following
discussion, $\hat{M}_{\mathcal{D}}$ denotes the maximum matching
of $\textsc{BG}(\mathcal{D})$ found by this way.

\begin{Example}
\begin{small}
See the economy given in Example 1. Let price vector
$\textbf{p}=(0,5,4,3,5)$ and $R$ be the rationing system such that
$R(i,a)=1$ for all $\langle i,a\rangle\in N\times X$. Then buyers'
constrained demands at $\textbf{p}$ and $R$ are:
$D_1(\textbf{p},R)=\{c,d\}$,
$D_2(\textbf{p},R)=D_3(\textbf{p},R)=\{c\}$,
$D_4(\textbf{p},R)=\{a\}$, $D_5(\textbf{p},R)=\{d\}$. Let
$\mathcal{D}=(D_i(\textbf{p},R))_{i\in N}$. Then
$\hat{M}_{\mathcal{D}}=\{\{1,c\},\{4,a\},\{5,d\}\}$.
\end{small}
\end{Example}

\section{Over-demanded Set of Items}
What can lead to non-existence of equilibrium allocations? This is a
key issue that we need to consider.

Given a demand situation $\mathcal{D}=(D_i)_{i\in N}$, a set of
real items $X'\subseteq X\setminus\{o\}$ is
\textit{over-demanded} in $\mathcal{D}$, if the number of buyers
who demand only items in $X'$ is strictly greater than the
number of items in $X'$, i.e., $|\{i\in N|D_i\subseteq
X'\}|>|X'|$; $X'$ is \textit{not under-demanded}, if the number of
buyers who demand some items in $X'$ is not less than the
number of items in $X'$, i.e., $|\{i\in N|D_i\cap
X'\neq\emptyset\}|\geq|X'|$. An over-demanded set $X'$ is
\textit{minimal} if no strict subset of $X'$ is over-demanded. We
can get Lemma 1 directly based on these definitions.
\begin{Lemma}
Let $X'\subseteq X\setminus\{o\}$ is over-demanded. Then for each
$a\in X'$, either there exists a minimal over-demanded set
$X''\subseteq X'$ s.t. $a\not\in X''$, or $a\in X''$ for every
minimal over-demanded set $X''\subseteq X'$.
\end{Lemma}

Theorem 1 answers the question proposed in the beginning of this section.

\begin{Theorem}
There exists an over-demanded set of items in
$\mathcal{D}=(D_i)_{i\in N}$ if and only if there does not exist
an equilibrium allocation.
\end{Theorem}
\textsc{Proof.} \begin{small} Sufficiency is obvious. Let us prove
necessity. Suppose there does not exist an equilibrium allocation.
Let $M=\hat{M}_{\mathcal{D}}$ and $N'=\{i\in N|o\not\in D_i\}$.
Then $|M|=|N\cap\bigcup_{e\in M}e|<|N'|$. Pick a buyer $i$ from
$N'\setminus N\cap\bigcup_{e\in M}e$. We construct a sequence
$\langle X_0,N_0\rangle,\langle X_1,N_1\rangle,\ldots$ as follow:
\begin{itemize}
\item $X_0=D_i$, $N_0=\{j\in N|(\exists a\in X_0)\{j,a\}\in M\}$;

\item $X_{k+1}=\bigcup_{j\in N_k}D_j$; and $N_{k+1}=\{j\in
N|(\exists a\in X_{k+1})\{j,a\}\in M\}$.
\end{itemize}
Pick any $k\geq 0$ and $a\in X_{k}$. Suppose there does not exist
$j\in N$ such that $\{j,a\}\in M$. Then there is an $M$-augmenting
path \cite{Schrijver} from $a$ to $i$, i.e., $M$ is not maximum,
contradicting the fact that $M$ is maximum. So for all $k\geq 0$
and $a\in X_{k}$, there exists $j\in N$ such that $\{j,a\}\in M$.
Consequently,
\begin{enumerate}
\item $X_k\subseteq X_{k+1}\subseteq X$, $N_k\subseteq
N_{k+1}\subseteq N$ for all $k\geq 0$;

\item if $X_{k+1}=X_{k}$ then $X_{k+l}=X_{k}$ and $N_{k+l}=N_{k}$
for all $k,l\geq 0$.
\end{enumerate}
So there must exist $K\geq 0$ s.t. $X_{0}\subset\ldots\subset
X_{K}=X_{K+1}=\ldots$. For each $b\in X_{K}$, $b$ is assigned to
only one buyer in $N_{K}$ at $\pi^M$. And for each $j\in N_{K}$,
$D_j\subseteq X_{K}$ and $j$ is assigned with only one item in
$X_{K}$ at $\pi^M$. So $|X_{K}|=|N_{K}|$. Consequently, $|\{i\in
N|D_i\subseteq
X_{K}\}|\geq|N_{K}\cup\{i\}|=|N_{K}|+1=|X_{K}|+1>|X_{K}|$. So
$X_{K}$ is an over-demanded set of items in
$\mathcal{D}$.$\quad\square$
\end{small}

\begin{figure}
\begin{center}
\begin{small}
\begin{tabular}{rl}
\hline\noalign{\smallskip}
1. & $\underline{\textbf{algorithm}}$ $\textsc{MODS}(\mathcal{D}=(D_i)_{i\in N},M=\hat{M}_{\mathcal{D}})$\\
2. & $\quad$pick $i$ from $\{i\in N|o\not\in D_i\}\setminus\bigcup_{e\in M}e$;\\
3. & $\quad X'':=D_i,X':=\emptyset$;\\
4. & $\quad\underline{\textbf{while}}(X''\neq\emptyset)$\\
5. & $\qquad N':=\{j\in N|(\exists a\in X'')\{j,a\}\in M\}$;\\
6. & $\qquad X':=X'\cup X''$, $X'':=\bigcup_{j\in N'}D_j\setminus X'$;\\
7. & $\quad X_{min}:=\emptyset,X'':=X'$;\\
8. & $\quad\underline{\textbf{for}}$ all $a\in X'$\\
9. & $\qquad X'':=X''\setminus\{a\}$;\\
10. & $\qquad N':=\{i\in N|D_i\subseteq X_{min}\cup X''\}$;\\
11. & $\qquad\mathcal{D'}:=(D_i)_{i\in N'}$, $k:=|\hat{M}_{\mathcal{D'}}|$;\\
12. & $\qquad\underline{\textbf{if}}$ $k=|N'|$\\
13. & $\qquad\quad X_{min}:=X_{min}\cup\{a\}$;\\
14. & $\quad\underline{\textbf{return}}$ $X_{min}$;\\
\noalign{\smallskip} \hline
\end{tabular}
\end{small}
\caption{\textsc{MODS} algorithm.}
\end{center}
\end{figure}

To find a minimal over-demanded set of items, we develop the
\textsc{MODS} algorithm shown in Figure 1. Given a demand
situation $\mathcal{D}$, and $\hat{M}_{\mathcal{D}}$ s.t.
$|\hat{M}_{\mathcal{D}}|<|\{i\in N|o\not\in D_i\}|$,
$\textsc{MODS}$ returns a minimal over-demanded set of items
$X_{min}$. The basic idea of $\textsc{MODS}$ is to generate an
over-demanded set $X'$ firstly (see lines 2-6 in Figure 1), and
then (according to Lemma 1) to find a minimal over-demanded set
$X_{min}\subseteq X'$ (see lines 7-14 in Figure 1).

The correctness of algorithm \textsc{MODS} is directly from Lemma
1 and the proof of Theorem 1. Let $BG(\mathcal{D})=\langle
\mathcal{V,E}\rangle$. Observe \textsc{MODS} and we can find the
following facts.
\begin{enumerate}
\item In order to generate an over-demanded set $X'$ (lines 4-6 in
Figure 1), \textsc{MODS} only visits edges in $\mathcal{E}$. For
each $e\in\mathcal{E}$, $e$ can be visited once at most.

\item $|X'|\leq|\hat{M}_{\mathcal{D}}|\leq\min(|N|,|X|)$, and
$BG(\mathcal{D'})\subseteq BG(\mathcal{D})$ (see line 11).

\end{enumerate}
According to $|\mathcal{E}|\leq|N||X|$, and that the complexity of
$\hat{M}_{\mathcal{D}}$ is in $O(|N||X|\min(|N|,|X|))$, the
overall complexity of
$\textsc{MODS}(\mathcal{D},\hat{M}_{\mathcal{D}})$ is in
$O(|N||X|(\min(|N|,|X|))^2)$.
\begin{Example}
\begin{small}
See $\mathcal{D}$ and $\hat{M}_{\mathcal{D}}$ described in Example
2. It is easy to find that $|\hat{M}_{\mathcal{D}}|<|\{i\in N|
o\not\in D_i\}|$. We apply $\textsc{MODS}$ algorithm to
$(\mathcal{D},\hat{M}_{\mathcal{D}})$. Firstly, an over-demanded
set $X'=\{c,d\}$ is found. And then a minimal over-demanded set
$X_{min}=\{c\}$ is found.
\end{small}
\end{Example}

\section{Mechanism for Resource Allocation under Price Rigidities}

In this section, we present a polynomial mechanism for resource
allocation under price rigidities (MAPR). Its basic idea is to
eliminate over-demanded sets of items by increasing the prices
of over-demanded items or rationing an over-demanded item
whose price has reached its upper bound.

\begin{center}
\textsc{MAPR}
\end{center}

\begin{itemize}
\item[\textbf{(1)}] The seller $\varphi$ announces the set $X$
of items to allocate, and sets
$\textbf{p}^0:=\underline{\textbf{p}},M^0:=\emptyset,N':=N$. Each
buyer $i\in N$ sets $R_i[a]:=1$ for all $a\in X$. Let $t:=0$.

\item[\textbf{(2)}] $\varphi$ sends $\textbf{p}^t$ and ``Report
your demand." to each $i\in N'$.

\item[\textbf{(3)}] Each $i\in N'$ computes and sends
$D_i$\footnote{$D_i=\{a\in X|R_i[a]=1\textrm{ and
}u_i(a)-\textbf{p}^t_a=\max\{u_i(b)-\textbf{p}^t_b|R_i[b]=1\}\}$}
to $\varphi$.

\item[\textbf{(4)}] $\varphi$ computes $N''=\{i\in
N'|D_i\cap\bigcup_{e\in M^t}e\neq\emptyset\}$. If $N''=\emptyset$
then go to step (6). $\varphi$ sends ``Sorry, items in
$D'_i=D_i\cap\bigcup_{e\in M^t}e$ have been sold. Please report
your new demand.'' to each $i\in N''$, and sets $N':=N''$.

\item[\textbf{(5)}] Each $i\in N'$ sets $R_i[a]:=0$ for all $a\in
D'_i$. Go to (3).

\item[\textbf{(6)}] Let $N^*=N\setminus\bigcup_{e\in M^t}e$ and
$\mathcal{D}^*=(D_i)_{i\in N^*}$. $\varphi$ computes
$\hat{M}_{\mathcal{D}^*}$. If $|\hat{M}_{\mathcal{D}^*}|=|\{i\in
N^*|o\not\in D_i\}|$ then go to step (9). $\varphi$ computes
$X_{min}=\textsc{MODS}(\mathcal{D}^*,\hat{M}_{\mathcal{D}^*})$.

\item[\textbf{(7)}] $\varphi$ computes $\overline{X}=\{a\in
X_{min}|\textbf{p}^t_a=\overline{\textbf{p}}_a\}$. If
$\overline{X}=\emptyset$ then: $\varphi$ sets $N':=N^*$,
$M^{t+1}:=M^t$, $\textbf{p}^{t+1}_a:=\textbf{p}^{t}_a+1$ for all
$a\in X_{min}$, and $\textbf{p}^{t+1}_a:=\textbf{p}^{t}_a$ for all
$a\in X\setminus X_{min}$. Let t:=t+1. Go to (2).

\item[\textbf{(8)}] $\varphi$ picks an item $a$ from
$\overline{X}$ and asks the buyers in $\{i\in N^*|a\in
D_i\subseteq X_{min}\}$ to draw lots for the right to buy $a$. Let
$i$ be the winning buyer. $\varphi$ sets
$M^{t+1}:=M^t\cup\{\{i,a\}\}$, $N':=N^*\setminus\{i\}$ and
$\textbf{p}^{t+1}:=\textbf{p}^{t}$. Let t:=t+1. Go to (2).

\item[\textbf{(9)}] $\varphi$ computes
$M^*:=M^t\cup\textsc{RM}((D_i)_{i\in
N},M^t,\textbf{p}^t,\underline{\textbf{p}})$ and then announces
$\textbf{p}^{t}$ and $\pi^{M^*}$ are the final price vector and
allocation. MAPR stops.
\end{itemize}

\begin{figure}
\begin{center}
\begin{small}
\begin{tabular}{rl}
\hline\noalign{\smallskip}
1. & $\underline{\textbf{algorithm}}$ $\textsc{RM}((D_i)_{i\in N},M,\textbf{p},\underline{\textbf{p}})$\\
2. & $\quad X':=\{a\in X\setminus\bigcup_{e\in M}e|\textbf{p}_a>\underline{\textbf{p}}_a\}$;\\
3. & $\quad N':=\{i\in N\setminus\bigcup_{e\in M}e|D_i\cap X'\neq\emptyset\}$;\\
4. & $\quad \mathcal{D'}:=(D_i\cap X')_{i\in N'},\; M':=\hat{M}_{\mathcal{D'}}$;\\
5. & $\quad N^*:=N\setminus\bigcup_{e\in M}e$, $\langle\mathcal{V,E}\rangle:=BG((D_i)_{i\in N^*})$;\\
6. & $\quad M'':=M'\cap\mathcal{E}$;\\
7. & $\quad \underline{\textbf{while}}(\textsc{Ma}(\langle\mathcal{V,E}\rangle,M'')\neq M'')$\\
8. & $\qquad M'':=\textsc{Ma}(\langle\mathcal{V,E}\rangle,M'')$;\\
9. & $\quad \underline{\textbf{return}}$ $M''\cup \{e\in M'|e\cap \bigcup_{e'\in M''}e'=\emptyset\}$;\\
\noalign{\smallskip} \hline
\end{tabular}
\end{small}
\caption{\textsc{RM} algorithm.}
\end{center}
\end{figure}

\cite{Talman}provides two dynamic procedures that produce constrained Walrasian equilibrium.
But it does not address the computation issues, and the third condition of constrained Walrasian equilibrium cannot be guaranteed either. In order to make sure that all the items whose prices exceed
their lower bound prices will be sold(the third criterion of constrained Walrasian equilibrium),
the $\textsc{RM}$
subroutine shown in Figure 2 is called in step 9. Given a demand
situation $\mathcal{D}=(D_i)_{i\in N}$, a partial matching $M$
consistent with $\mathcal{D}$, the current price vector
$\textbf{p}$, and the lower bound price vector
$\underline{\textbf{p}}$, $\textsc{RM}$ returns a matching $M'$
such that \textbf{(1)} $\pi^{M\cup M'}$ is an equilibrium
allocation, \textbf{(2)} $M\cap M'=\emptyset$, and \textbf{(3)}
$\{a\in X\setminus\bigcup_{e\in
M}e|\textbf{p}_a>\underline{\textbf{p}}_a\}\subseteq\bigcup_{e\in
M'}e$.

Observe MAPR and RM subroutine. We can find that:
\begin{itemize}
\item computation of each step is polynomial in $|N|$ and $|X|$;

\item for each $t\geq 0$, the number of the loops consisting of
steps 3-5 is not more than $|X|$; and

\item the number of the loops consisting of steps 2-8 is not more
than $\sum\limits_{a\in
X}(\overline{\textbf{p}}_a-\underline{\textbf{p}}_a)$.
\end{itemize}
Consequently, MAPR always terminates and is polynomial in $|N|$,
$|X|$, and $\sum\limits_{a\in
X}(\overline{\textbf{p}}_a-\underline{\textbf{p}}_a)$.

In order to prove the correctness of MAPR and RM, we will first
give some definitions and provide three lemmas, then we will prove
that MAPR can lead to a constrained Walrasian equilibrium with the
help of these three lemmas. In the following discussion, we
suppose that MAPR terminates at some time $T\geq 0$;
$\textbf{p}^t$, $M^t$, $R^t$ ($R^t(i,a)=R_i[a]$ for all $\langle
i,a\rangle\in N\times X$, where $R_i$ is the vector kept by buyer
$i$ at time $t$), and $(D_i^t)_{i\in N}$ denote the price vector,
partial matching that has been made so far, rationing system, and
demand situation at time $0\leq t\leq T$, respectively. Let
$X^t=\{a\in X\setminus\bigcup_{e\in
M^t}e|\textbf{p}^t_a>\underline{\textbf{p}}_a\}$ and $N^t=\{i\in
N\setminus\bigcup_{e\in M^t}e| D^t_i\cap X^t\neq\emptyset\}$.

Now we introduce three auxiliary lemmas (in which
$\mathcal{D}=(D_i)_{i\in N}$ denotes a demand situation). These
three lemmas are closely connected. The proof of Lemma 4 is based
on Lemma 2 and Lemma 3, and the proof of Theorem 2 is based on the
these three lemmas. Lemma 2 states that, each nonempty subset of a
minimal over-demanded set of items is not under-demanded.
\begin{Lemma}
Let $X'$ be a minimal over-demanded set of items. Then for
each $\emptyset\subset X''\subseteq X'$, $|\{i\in N|D_i\cap
X''\neq\emptyset\textrm{ and }D_i\subseteq X'\}|>|X''|$.
\end{Lemma}
The proof of Lemma 2 is not very hard, and comes from using the
reduction to absurdity.

Lemma 3 states that, the cardinality of a maximum matching is not
less than the cardinality of a set of real items if each
subset of the set is not under-demanded.

\begin{Lemma}
Let $X'\subseteq X\setminus\{o\}$ and $|\{i\in N|D_i\cap
X''\neq\emptyset\}|\geq|X''|$ for each $X''\subseteq X'$. If $M$
is a maximum matching of $BG((D_i\setminus\{o\})_{i\in N})$, then
$|M|\geq|X'|$.
\end{Lemma}
The proof of Lemma 3 is similar to that of Theorem 1. Due to lack
of space, it is omitted.

Lemma 4 states that, all the items in $X^t$ can be sold. The proof of Lemma 4 is based on Lemma 2 and Lemma 3.
\begin{Lemma}
Let $\mathcal{D}^t=(D^t_i\cap X^t)_{i\in N^t}$. Then
$|\hat{M}_{\mathcal{D}^t}|=|X^t|$ for each $0\leq t\leq T$.
\end{Lemma}
\textsc{Proof.} \begin{small}We first prove that $|\{i\in
N^t|D^t_i\cap X'\neq\emptyset\}|\geq|X'|$ for each
$\emptyset\subset X'\subseteq X^t$ and $0\leq t\leq T$:
\begin{enumerate}
\item It holds at $t=0$ because $X^0=\emptyset$.

\item Suppose MAPR does not stop at $\hat{t}\geq 0$ and $|\{i\in
N^t|D^t_i\cap X'\neq\emptyset\}|\geq|X'|$ for each
$\emptyset\subset X'\subseteq X^t$ and $0\leq t\leq \hat{t}$.

\item Then $X_{min}\neq\emptyset$ and $\overline{X}$ are computed
at time $\hat{t}$ and steps 6-7 of MAPR. Pick any
$\emptyset\subset X'\subseteq X^{\hat{t}+1}$. Let $N_1=\{i\in
N^{\hat{t}}|D_i^{\hat{t}}\subseteq X_{min}\textrm{ and
}D_i^{\hat{t}}\cap X'\neq\emptyset\}$ and $N_2=\{i\in
N^{\hat{t}}|D_i^{\hat{t}}\cap (X'\setminus
X_{min})\neq\emptyset\}$. There are two possibilities:
\begin{itemize}
\item[Case I]: $\overline{X}=\emptyset$. So
$X^{\hat{t}+1}=X^{\hat{t}}\cup X_{min}$. According to Lemma 2 and
item 2, we have $|N_1|>|X'\cap X_{min}|$ and
$|N_2|\geq|X'\setminus X_{min}|$. It is easy to find that
$D_i^{\hat{t}+1}\cap X'\neq\emptyset$ for each $i\in N_1\cup
N_2\subseteq N^{\hat{t}+1}$ and $N_1\cap N_2=\emptyset$. So
$|\{i\in N^{\hat{t}+1}|D^{\hat{t}+1}_i\cap
X'\neq\emptyset\}|\geq|N_1\cup N_2|=|N_1|+|N_2|>|X'\cap
X_{min}|+|X'\setminus X_{min}|=|X'|$.

\item[Case II]: $\overline{X}\neq\emptyset$ and some $a\in
\overline{X}$ is assigned to some buyer $j$ such that $a\in
D_j^{\hat{t}}\subseteq X_{min}$. So
$X^{\hat{t}+1}=X^{\hat{t}}\setminus\{a\}$. According to Lemma 2
and item 2, we have $|N_1|>|X'\cap X_{min}|$ and
$|N_2|\geq|X'\setminus X_{min}|$. It is easy to find that
$D_i^{\hat{t}+1}\cap X'\neq\emptyset$ for each $i\in
(N_1\setminus\{j\})\cup N_2\subseteq N^{\hat{t}+1}$ and $N_1\cap
N_2=\emptyset$. Consequently, $|\{i\in
N^{\hat{t}+1}|D^{\hat{t}+1}_i\cap
X'\neq\emptyset\}|\geq|(N_1\setminus\{j\})\cup
N_2|\geq|N_1|-1+|N_2|\geq|X'\cap X_{min}|+|X'\setminus
X_{min}|=|X'|$.
\end{itemize}
Consequently, $|\{i\in N^{\hat{t}+1}|D^{\hat{t}+1}_i\cap
X'\neq\emptyset\}|\geq|X'|$.
\end{enumerate}
According to items 1--3, $|\{i\in N^t|D^t_i\cap
X'\neq\emptyset\}|\geq|X'|$ for each $X'\subseteq X^t$ and $0\leq
t\leq T$. It is easy to find that
$|\hat{M}_{\mathcal{D}^t}|\leq|X^t|$ for each $0\leq t\leq T$.
According to Lemma 3, we have $|\hat{M}_{\mathcal{D}^t}|\geq|X^t|$
for each $0\leq t\leq T$. So $|\hat{M}_{\mathcal{D}^t}|=|X^t|$ for
each $0\leq t\leq T$.$\quad\square$
\end{small}

Now we are ready to establish the following correctness theorem
for MAPR (and RM subroutine).
\begin{Theorem}
$\langle\textbf{p}^T,R^T,\pi^{M^T}\rangle$ found by MAPR, is a
constrained Walrasian equilibrium.
\end{Theorem}
\textsc{Proof.} (Sketch)
\begin{small}$\langle\textbf{p}^T,R^T,\pi^{M^T}\rangle$ is a
constrained Walrasian equilibrium iff it satisfies the five
conditions shown in page 2.
\begin{enumerate}
\item It is easy to find that conditions (1), (4), and (5) are
satisfied by $\langle\textbf{p}^T,R^T,\pi^{M^T}\rangle$.

\item For each buyer $i$ and the item assigned to her
$a=\pi^{M^T}(i)$, there are two possibilities: Case I (step (8)),
$i$ is the winner of a lottery on item $a$ at some time
$T'\leq T$, and Case II (step (6) and (9)), $a$ is assigned to $i$
at time $T$.
\begin{enumerate}
\item In case I, $a\in D_i(\textbf{p}^{T'},R^{T'})$. So
$u_i(a)-\textbf{p}^{T'}_a\geq u_i(b)-\textbf{p}^{T'}_b$ for all
$b\in \{b\in X|R^{T'}(i,b)=1\}$. Because
$R^{T'}(i,a)=R^{T}(i,a)=1$, $\textbf{p}^{T'}_a=\textbf{p}^{T}_a$,
$R^{T'}(i,b)\geq R^{T}(i,b)$ and
$\textbf{p}^{T'}_b\leq\textbf{p}^{T}_b$ for all $b\in X$,
$u_i(a)-\textbf{p}^{T}_a\geq u_i(b)-\textbf{p}^{T}_b$ for all
$b\in\{b\in X|R^{T}(i,b)=1\}$. So $a\in
D_i(\textbf{p}^{T},R^{T})$.

\item In case II, according to the definition of $\pi^{M^T}$ (see
RM subroutine and steps (6)--(9)), we have $a\in
D_i(\textbf{p}^{T},R^{T})$.
\end{enumerate}
Consequently, $\pi^{M^T}$ is an equilibrium allocation.

\item According to Lemma 4, all the items in $X^T$ are sold.
Consequently, $\textbf{p}^T_a=\underline{\textbf{p}}_a$ for each
$a\in\{b\in X|(\forall i\in N)\pi^{M^T}(i)\neq b\}$. The
correctness of RM subroutine can derive from item 2 and item 3
directly.
\end{enumerate}
So $\langle\textbf{p}^T,R^T,\pi^{M^T}\rangle$ is a constrained
Walrasian equilibrium.$\quad\square$
\end{small}
\begin{Example}
\begin{small}
See Example 1. Apply MAPR to $\langle
E,\underline{\textbf{p}},\overline{\textbf{p}}\rangle$. The
demands, price vectors, rationing system and other relevant data
generated by MAPR are illustrated in Table 2, where $U_i$, $D_i$,
$X'$, $N'$, and $X_{min}$ denote $\{a\in X|R^t(i,a)=0\}$,
$D_i(\textbf{p}^t,R^t)$, $X\cap\bigcup_{e\in M^t}e$,
$N\cap\bigcup_{e\in M^t}e$, and the value of $X_{min}$ computed by
the seller at step (6) and time $t$.

At $t=3$, the price of $c$ has reached its upper bound 4. The
seller assigns randomly $c$ to buyer $2$ or buyer $3$. So there
are two different possible histories of resource allocation from
$t=3$. Along the history of $t=4.1;5.1;6.1$, MAPR finds
$\langle\textbf{p}^{6.1},R^{6.1},\pi^{M^{6.1}}\rangle$, where
$\pi^{M^{6.1}}(1)=o$, $\pi^{M^{6.1}}(2)=c$, $\pi^{M^{6.1}}(3)=b$,
$\pi^{M^{6.1}}(4)=a$, and $\pi^{M^{6.1}}(5)=d$. Along the history
of $t=4.2;5.2;6.2$, MAPR finds
$\langle\textbf{p}^{6.2},R^{6.2},\pi^{M^{6.2}}\rangle$, where
$\pi^{M^{6.2}}(1)=o$, $\pi^{M^{6.2}}(2)=b$, $\pi^{M^{6.2}}(3)=c$,
$\pi^{M^{6.2}}(4)=a$, and $\pi^{M^{6.2}}(5)=d$.
\end{small}
\end{Example}
\begin{table*}
\caption{Data Generated by MAPR}
\begin{center}
\begin{small}
\begin{tabular}{c|c|c|c|c|c|c|c|c|c|c|c|c|c|c|c|c|c|c}
\hline
t & $\textbf{p}^t_o$ & $\textbf{p}^t_a$ & $\textbf{p}^t_b$ & $\textbf{p}^t_c$ & $\textbf{p}^t_d$ & $X_{min}$ & $U_1$ & $U_2$ & $U_3$ & $U_4$ & $U_5$ & $N'$ & $D_1$ & $D_2$ & $D_3$ & $D_4$ & $D_5$ & $X'$\\
\hline
0 & 0 & 5 & 4 & 1 & 5 & $\{c\}$ & $\emptyset$ & $\emptyset$ & $\emptyset$ & $\emptyset$ & $\emptyset$ & $\emptyset$ & $\{c\}$ & $\{c\}$ & $\{c\}$ & $\{a\}$ & $\{d\}$ & $\emptyset$\\
1 & 0 & 5 & 4 & 2 & 5 & $\{c\}$ & $\emptyset$ & $\emptyset$ & $\emptyset$ & $\emptyset$ & $\emptyset$ & $\emptyset$ & $\{c\}$ & $\{c\}$ & $\{c\}$ & $\{a\}$ & $\{d\}$ & $\emptyset$\\
2 & 0 & 5 & 4 & 3 & 5 & $\{c\}$  & $\emptyset$ & $\emptyset$ & $\emptyset$ & $\emptyset$ & $\emptyset$ & $\emptyset$ & $\{c,d\}$ & $\{c\}$ & $\{c\}$ & $\{a\}$ & $\{d\}$ & $\emptyset$\\
3 & 0 & 5 & 4 & 4 & 5 & $\{c\}$ & $\emptyset$ & $\emptyset$ & $\emptyset$ & $\emptyset$ & $\emptyset$ & $\emptyset$ & $\{d\}$ & $\{c\}$ & $\{c\}$ & $\{a\}$ & $\{d\}$ & $\emptyset$\\
4.1 & 0 & 5 & 4 & 4 & 5 & $\{d\}$ & $\emptyset$ & $\emptyset$ & $\{c\}$ & $\emptyset$ & $\emptyset$ & $\{2\}$ & $\{d\}$ & & $\{d\}$ & $\{a\}$ & $\{d\}$ & $\{c\}$\\
5.1 & 0 & 5 & 4 & 4 & 6 & $\{d\}$ & $\{c\}$ & $\emptyset$ & $\{c\}$ & $\emptyset$ & $\emptyset$ & $\{2\}$ & $\{d\}$ & & $\{b,d\}$ & $\{a\}$ & $\{d\}$ & $\{c\}$\\
6.1 & 0 & 5 & 4 & 4 & 7 & $\emptyset$ & $\{c\}$ & $\emptyset$ & $\{c\}$ & $\emptyset$ & $\emptyset$ & $\{2\}$ & $\{o,d\}$ & & $\{b\}$ & $\{a\}$ & $\{d\}$ & $\{c\}$\\
4.2 & 0 & 5 & 4 & 4 & 5 & $\{d\}$ & $\emptyset$ & $\{c\}$ & $\emptyset$ & $\emptyset$ & $\emptyset$ & $\{3\}$ & $\{d\}$ & $\{a,b\}$ & & $\{a\}$ & $\{d\}$ & $\{c\}$\\
5.2 & 0 & 5 & 4 & 4 & 6 & $\{d\}$  & $\{c\}$ & $\{c\}$ & $\emptyset$ & $\emptyset$ & $\emptyset$ & $\{3\}$ & $\{d\}$ & $\{a,b\}$ & & $\{a\}$ & $\{d\}$ & $\{c\}$\\
6.2 & 0 & 5 & 4 & 4 & 7 & $\emptyset$  & $\{c\}$ & $\{c\}$ & $\emptyset$ & $\emptyset$ & $\emptyset$ & $\{3\}$  & $\{o,d\}$ & $\{a,b\}$ & & $\{a\}$ & $\{d\}$ & $\{c\}$\\
\hline
\end{tabular}
\end{small}
\end{center}
\end{table*}

\section{Expected profits, Expected Prices, and Strategical Issues}

Since the history of MAPR is nondeterministic, we need to
introduce concepts of buyers' \emph{expected profits} and
items' \emph{expected prices}. Let $R_*^t$ be a rationing
system s.t. $R_*^t(i,a)=1$ if $\{i,a\}\in M^t$ or
$a\not\in\bigcup_{e\in M^t}e$, and 0 otherwise. Because we can
induce $M^t$ from $R_*^t$. So $M^t$ can be written as $M^{R_*^t}$.
We say $\langle \textbf{p}^t,R_*^t\rangle$ is an allocation
situation. Assume that the computation of $\textsc{MODS}$
algorithm and the selection of items in step (8) are
deterministic, all the lots happening in MAPR are fair
\footnote{Suppose there are $k$ buyers drawing lots for the right
to buy item $a$. Then the lot is fair if each one of these
buyers has $1/k$ chance of winning the lot.}. Then buyer $i$'s
expected profit and item $a$'s expected price on $\langle
\textbf{p},R\rangle$ (i.e., $u^*_i(\textbf{p},R)$ and
$\textbf{p}^*_a(\textbf{p},R)$) are:
\begin{small}
\begin{displaymath}
u^*_i(\textbf{p},R)=\left\{\begin{array}{ll} V_{i}(\textbf{p},R) & \textrm{if}\:X_{min}=\emptyset\\
u^*_{i}(\textbf{p}',R) &
\textrm{if}\:\overline{X}=\emptyset\\
\frac{\sum_{i'\in N'}u^*_{i}(\textbf{p},R_{i'})}{|N'|} &
\textrm{otherwise}
\end{array} \right.
\end{displaymath}
\begin{displaymath}
\textbf{p}_a^*(\textbf{p},R)=\left\{\begin{array}{ll} \textbf{p}_a & \textrm{if}\:X_{min}=\emptyset\\
\textbf{p}_a^*(\textbf{p}',R) &
\textrm{if}\:\overline{X}=\emptyset\\
\frac{\sum_{i'\in N'}\textbf{p}_a^*(\textbf{p},R_{i'})}{|N'|} &
\textrm{otherwise}
\end{array} \right.
\end{displaymath}
where (let $\mathcal{D}=(D_i(\textbf{p},R))_{i\in N}$):
\begin{itemize} \item $X_{min}=\emptyset$ if $|\hat{M}_{\mathcal{D}}|=|\{i\in N|o\not\in D_i(\textbf{p},R)\}|$, and $\textsc{MODS}(\mathcal{D},\hat{M}_{\mathcal{D}})$
otherwise; $\overline{X}=\{a\in
X_{min}|\textbf{p}_a=\overline{\textbf{p}}_a\}$;

\item $\textbf{p}'_{a}=\textbf{p}_{a}$ for all $a\not\in X_{min}$
and $\textbf{p}'_{a}=\textbf{p}_{a}+1$ for all $a\in X_{min}$;

\item $b\in \overline{X}$ is the item selected by the
seller in step (8);

\item $N'=\{i\in N|b\in D_i(\textbf{p},R)\subseteq X_{min}\}$;

\item for all $\langle i,a\rangle\in N\times X$:
$R_{i'}(i,a)=R(i,a)$ if $a\neq b$; $R_{i'}(i,b)=0$ if $i\neq i'$;
and $R_{i'}(i',b)=1$.
\end{itemize}
\end{small}
In fact, $u^*_i(\textbf{p},R)$ and $\textbf{p}^*_a(\textbf{p},R)$
can be computed by developing a search tree: each node is an
allocation situation, and is expanded (if $X_{min}\neq\emptyset$)
into (i) one single branch if $\overline{X}=\emptyset$, and (ii)
$|N'|$ branches otherwise. See Table 1 and Table 2. We can find
that
$u^*_1(\textbf{p}^0,R^0_*)=0.5*u^*_1(\textbf{p}^{6.1},R^{6.1}_*)+0.5*u^*_1(\textbf{p}^{6.2},R^{6.2}_*)=0$,
$u^*_3(\textbf{p}^0,R^0_*)=0.5*u^*_3(\textbf{p}^{6.1},R^{6.1}_*)+0.5*u^*_3(\textbf{p}^{6.2},R^{6.2}_*)=2.5$,
$\textbf{p}_a^*(\textbf{p}^0,R^0_*)=0.5*\textbf{p}_a^*(\textbf{p}^{6.1},R^{6.1}_*)+0.5*\textbf{p}_a^*(\textbf{p}^{6.2},R^{6.2}_*)=5$.

As most collective decision mechanisms, MAPR is generally not
\emph{strategyproof} (in the sense of expected profit). For
instance, see Example 4. If buyer 1 reports her demands sincerely,
then her expected profit is 0. However, if 1 knows other buyers'
valuations and reports strategically, then she reports $\{c\}$
from $t=0$ to $t=3$ (i.e., as if her valuation to item $c$ is
not less than 7), then reports sincerely, then her expected
profit changes to 1/3, which makes her better off.

Now we are interested in two questions: (1) is MAPR strategyproof
for some restricted domains? (2) when it is not, how hard is it
for an buyer who knows the valuations of the others to compute an
optimal strategy?

First we define reporting strategies and manipulation problems
formally. Without loss of generality, let 1 be the manipulator.
Note that not every sequence of 1's demands is reasonable. For
instance, see Example 4 and Table 2. The seller can detect 1's
manipulation if 1 reports $\{c\}$, $\{c\}$, $\{c,d\}$, and $\{c\}$
at $t=0$, 1, 2, and 3, respectively, because there is no value
function $u$ s.t.
$u(c)-\textbf{p}^2_c=u(c)-3=u(d)-5=u(d)-\textbf{p}^2_d=u(d)-\textbf{p}^3_d<u(c)-\textbf{p}^3_c=u(c)-4$.
A \emph{strategy} for buyer 1 is a value function $u: X
\rightarrow \mathbb{Z}_+$ with $u(o)=0$. So 1 can safely
manipulate the process of MAPR when she reports her demands
according to $u$ completely (as if $u$ is her true value
function). A \emph{manipulation problem} M (for buyer 1) is a
5-tuple $\langle N,X,\{u_i\}_{i\in N}, \underline{\textbf{p}},
\overline{\textbf{p}}\rangle$ where $\langle N,X,\{u_i\}_{i\in
N}\rangle$ is an economy, $\underline{\textbf{p}}$ and
$\overline{\textbf{p}}$ are the lower and upper bound price
vectors on $X$, respectively. A strategy for M is \emph{optimal}
if 1 can not strictly increase her expected profit by reporting
her demands according to any other strategy.

Now, back to question (1): we show that the answer is positive
when there are two buyers.

\begin{Theorem}
Let $M=\langle N,X,\{u_i\}_{i\in N}, \underline{\textbf{p}},
\overline{\textbf{p}}\rangle$ be a manipulation problem s.t.
$N=\{1,2\}$. Then $u_1$ is optimal for $M$.
\end{Theorem}
\textsc{Proof.}
\begin{small}
Suppose that if 1 reports sincerely, then her expected profit is
$\Delta$. Let $D_1$ and $D_2$ be 1 and 2's true demands at
$\underline{\textbf{p}}$ and $R$ respectively, where $R(i,a)=1$
for each $i\in N$ and $a\in X$.

Obviously, if $D_1\cup D_2=\{o\}$ or $|D_1\cup D_2|\geq 2$ (i.e.,
$X_{min}=\emptyset$ at $t=0$) then $\Delta=\max_{a\in
X}(u_1(a)-\underline{\textbf{p}}_a)$, which is the best possible
outcome for 1. So $u_1$ is optimal in these cases.

Now, suppose $D_1=D_2=\{a\}$ s.t. $a\neq o$. Pick any strategy
$u'$. Let $k=\overline{\textbf{p}}_a-\underline{\textbf{p}}_a$,
$k_i=u_i(a)-\underline{\textbf{p}}_a-\max_{b\in
X\setminus\{a\}}(u_i(b)-\underline{\textbf{p}}_b)$, $b_i\in
X\setminus\{a\}$ s.t. $u_i(b_i)-\underline{\textbf{p}}_{b_i}
=u_i(a)-\underline{\textbf{p}}_a-k_i$, and
$\hat{k}=\min(k,k_1-1,k_2-1)$. Then if 1 applies strategy $u_1$,
then she will report $D_1$ from $t=0$ to $t=\hat{k}$ and:
\begin{enumerate}
\item if $\hat{k}=k$, then $\Delta=0.5\ast
(u_1(a)-\underline{\textbf{p}}_a-k)+0.5\ast(u_1(b_1)-\underline{\textbf{p}}_{b_1})$=$u_1(b_1)-\underline{\textbf{p}}_{b_1}+0.5\ast(k_1-k)>u_1(b_1)-\underline{\textbf{p}}_{b_1}$.
If 1 applies $u'$ instead, then her expected profit will not be
better than $u_1(b_1)-\underline{\textbf{p}}_{b_1}<\Delta$ if $
u'(a)-\underline{\textbf{p}}_a-\max_{b\in
X\setminus\{a\}}(u'(b)-\underline{\textbf{p}}_b)\leq k$, and will
not be better than $\Delta$ otherwise.

\item if $k>\hat{k}=k_1-1$, then
$\Delta=u_1(b_1)-\underline{\textbf{p}}_{b_1}$. Because 2 can
insist on $\{a\}$ to $t=\min(k,k_2-1)\geq k_1-1$, 1's expected
profit can not be better than $\Delta$.

\item if $k>\hat{k}=k_2-1$, then
$\Delta=u_1(a)-\underline{\textbf{p}}_{a}-k_2\geq
u_1(a)-\underline{\textbf{p}}_{a}-k_1=u_1(b_1)-\underline{\textbf{p}}_{b_1}$.
Because 2 can insist on $\{a\}$ to $t=k_2-1$, 1's expected profit
can not be better than $\Delta$.
\end{enumerate}
To sum up, in all cases, 1 can not strictly increase her expected
profit by applying strategy $u'$. So $u_1$ is optimal for $M$.
$\quad\square$
\end{small}

For the cases where there are more than two buyers, we conjecture
that the manipulation problem is NP-hard, but we could not find a
proof.

\section{Conclusion}
We have presented a decentralized protocol for allocating
indivisible resources under price rigidities, and proved formally
that it can discover constrained Walrasian equilibria in
polynomial time. We also have studied the protocol from the points
of computation of buyers' expected profits and items'
expected prices, and discussed the manipulation (by one buyer)
problem in the sense of buyer's expected profit. There are
several directions for future work. One direction would be to
prove the conjecture about the complexity of manipulation (in the
sense of expected profits) by one buyer. Another direction would
be to study manipulation (in the sense of expected prices) by one or
more buyers (whose manipulation motivation is not to buy some
resources but to put up the prices of some resources).
Furthermore, we plan to study the problems of allocating divisible
resources \cite{Brams5} and sharable resources \cite{Airiau} under
prices rigidities.

\section*{Acknowledgments}
This work is supported by the National Natural Science Foundation
of China under grant No.61070232, No.61105039, No.61272295, and
the Fundamental Research Funds for the Central Universities (Issue
Number WK0110000026).

\bibliographystyle{named}
\bibliography{ijcai13}

\end{document}